\documentclass[prb,superscriptaddress,floatfix,twocolumn,showpacs,amsfonts,amsmath,amssymb]{revtex4}
\usepackage{graphicx} % include figure files
\usepackage{dcolumn} % align table columns on decimal point
\usepackage{bm} % bold math
\usepackage{color}
\newcommand{\cred}{\color{red}}
\usepackage{tikz}

\begin{document}

\title{Interference effects and Huygens' principle
in transverse magnetic focusing of electrons and holes}

\author{Samuel Bladwell}
\affiliation{School of Physics, University of New South Wales, Sydney 2052, Australia}
%\author{Alexander I. Milstein}
%\affiliation{Budker Institute of Nuclear Physics, 630090 Novosibirsk, Russia}
\author{Oleg P. Sushkov}
\affiliation{School of Physics, University of New South Wales, Sydney 2052, Australia}

\begin{abstract}

Interference effects form a fundamental pillar of quantum mechanics. In this paper, we examine the interference
in spin-orbit coupled transverse magnetic focusing, where a weak magnetic field is used to focus charge carries over
mesoscopic scales. We determine a semi-classical form for the Green's function in a weak magnetic field, for the case
of both spin-less and spin-orbit coupled charge carriers. The obtained forms for the Greens' function are independent of
particle dispersion and are thus applicable to a wide variety of systems.

\end{abstract}

\pacs{72.25.Dc, 71.70.Ej, 73.23.Ad  }% PACS, the Physics and Astronomy
\maketitle
%\tableofcontents

\section{Introduction}

\begin{figure}[h!]
     {\includegraphics[width=0.25\textwidth]{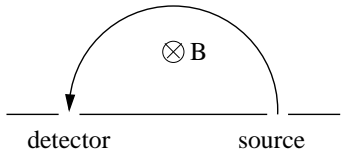}}
	\caption{Focusing in a weak magnetic field.}
\label{Fig1}
\end{figure}

Transverse magnetic focusing (TMF) is an experimental technique involving the focusing of charge carriers
from a source to a detector over a scale of microns via a weak magnetic field. 
This experimental technique is the direct translation of charge mass spectroscopy to the
solid state. A typical setup is presented in Fig.~\ref{Fig1}. 
The technique was proposed by Sharvin for studies of the Fermi surface in metals~\cite{Sharvin1965}.
See also Refs.~\cite{Sharvin1965a,Tsoi1974} and a review paper~\cite{Tsoi1999}.
More recently the technique was applied to two-dimensional (2D) semiconductor 
heterostructures~\cite{Vanhouten,Rokhinson2004,Rokhinson2006,Reynoso2007,Li2012}
and to graphene~\cite{Taychatanapat2013}.
In the case of semiconductors usually quantum point contacts (QPCs)  are used as the source and the detector. 
TMF is of a special importance for hole doped 2D semiconductor heterostructures due to the strong spin-orbit
interaction; hole trajectories for distinct spin polarizations are different resulting
in a double peak in the magnetic focusing spectrum~\cite{Rokhinson2004,Usaj2004,Zulicke2007,Schliemann2008,Kormanyos2010,
Bladwell}.
Hence, ``double" focusing is a method for studying the spin-orbit interaction, complementary to Shubnikov
de Haas oscillations.

\begin{figure}[h!]
     {\includegraphics[width=0.25\textwidth]{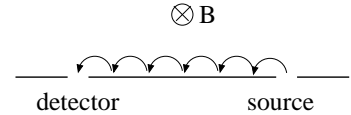}}
	\caption{The edge state in a  magnetic field.}
\label{Fig2}
\end{figure}

Usually TMF is considered in a classical regime, 
where after the spin-splitting of the trajectory is taken into account the analysis of the orbital dynamics is purely classical. 
Some orbital quantum interference effects 
have been considered for edge states with multiple specular reflections off the boundary, 
see Fig.~\ref{Fig2}, Refs.~\cite{Vanhouten,Usaj2004,Kormanyos2010,Bladwell}.
While this classical approximation is well justified for metals, the approximation is less appropriate for semiconductors,
since the typical Fermi wavelength in semiconductor heterostructures, $\lambda_F \sim 50-100$nm, is not small compared to the typical 
geometric size, $L\sim 1-2\mu m$. As a result interference effects are significant, and they can substantially alter the double 
focusing picture.

In this paper we consider orbital interference effects in the TMF geometry shown in Fig.~\ref{Fig1}.
The developed technique is quite general, results are naturally articulated in terms of the
 Huygens' principle.  The wave function at the position of the detector ${\bm R}$ is
expressed in terms of the wave function at the source ${\bm r}$,
\begin{equation}
\label{HF}
\psi({\bm R})=\int_{\Phi}K({\bm R}-{\bm r})\psi({\bm r}) dr
\end{equation}
The integration is performed over the wavefront $\Phi$ of the wave emitted from the source.
In this work we find the kernel $K({\bm r})$. Since we consider 2D heterostructures the kernel $K$ has
dimension $1/[r]$, generally $K$ is a matrix in spin space.
Analogous to optics the kernel $K$ is proportional to the Green's function.
Specifically, we consider the three following situations;
(i) TMF in the absence of spin orbit interactions (SOI), (ii) TMF with linear in momentum Rashba SOI,
 and (iii) TMF with cubic in momentum Rashba SOI. Addressing issues (ii) and (iii) we assume that the
SOI interaction is sufficiently strong to provide adiabatic spin dynamics, this is the limit of practical
importance for double focusing. The adiabatic transport of spin means that spin influences the interference 
picture only via the Berry phase.
In Section \ref{nospinorbit} we consider the interference picture without any SOI, and 
in Section \ref{withspinorbit} we consider the interference with account of Rashba SOI.
Finally Section \ref{disc} presents discussion and some specific numerical examples.

\section{Quantum interference in TMF without SOI}
\label{nospinorbit}
We start from semiclassical analysis which gives a physical insight in the problem
and is similar to that of Ref.~\cite{Vanhouten}.
Let a pointlike source be located at the coordinate origin and let
$\lambda$ be the wavelength and $r_c$ be the cyclotron radius.
Then at a distance $r$ within the range $\lambda \ll r \ll r_c$ the magnetic field is irrelevant, and 
the form of the Huygens kernel is
\begin{equation}
\label{H1}
K(r)=a\frac{e^{ikr}}{\sqrt{r}} \ ,
\end{equation}
where $a$ is a constant. We remind that in the 3D case 
$K(r)=a\frac{e^{ikr}}{r}$, where $a=\frac{k}{2\pi i}$ is independent of the particle dispersion~\cite{LL2}.
The dimension of the 3D Huygens' Kernel is $1/r^2$.
Applying  the method of Ref.~\cite{LL2} to the 2D case we find that the coefficient in Eq.(\ref{H1}) is
\begin{equation}
\label{aa}
a = \sqrt{\frac{k}{2\pi}}e^{-i\pi/4} \ . 
\end{equation}
Analogous to the 3D case, the coefficient is independent of the particle dispersion.

\begin{figure}[h!]
     {\includegraphics[width=0.49\textwidth]{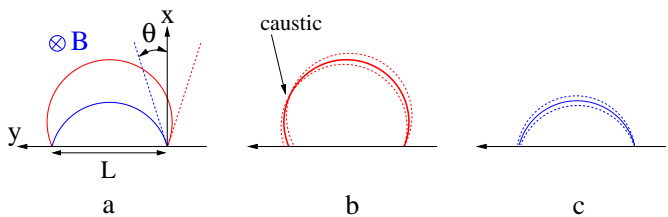}}
	\caption{Panel a: Two arc trajectories in magnetic field connecting the source and the detector
are shown by solid blue and red lines. Blue and red dashed lines are tangential to the corresponding
trajectories at the point of injection.
Panel b illustrates that the longer trajectory (red) has a caustic: trajectories with slightly different
injection angles always intersect.
Panel c illustrates that the shorter trajectory (blue) does not  have a caustic: trajectories with slightly different
injection angles do not intersect.
}
\label{Fig3}
\end{figure}

Now consider a point-like detector located
at $(0,L)$; see Panel a in Fig.~\ref{Fig3}.
In this geometry the magnetic field is important.
A peculiarity of TMF is that classically there are always two trajectories connecting the source and the detector,
so the picture is not like that in Fig.~\ref{Fig1}, the correct picture is shown in Fig.~\ref{Fig3}a.
In this figure we exaggerate the difference between the ``blue'' and  ``red'' trajectories. 
Each trajectory is an arc of the cyclotron circle with radius $r_c$.
The injection angle $\theta$ is related to the source - detector distance $L$,
\begin{equation}
\label{rel1}
L= 2 r_c \cos \theta \ .
 \end{equation}
This equation has two solutions, positive $\theta$ and negative $\theta$ corresponding to the blue and red
trajectories in Fig.~\ref{Fig3}a respectively.

The action along a classical trajectory is $S=\int{\bm p}\cdot d{\bm l}$, where ${\bm p}$ is canonical momentum.
The action is not gauge invariant and hereafter we choose the vector potential as
 \begin{equation}
\label{gg}
{\bm A}=(0,-Bx,0)\ ,
\end{equation}
where $B$ is the magnetic field.
Hereafter we set $\hbar=1$.
The relation between the kinematic momentum ${\bm k}$ and the canonical momentum ${\bm p}$ is
\begin{equation}
\label{kp} 
{\bm k}={\bm p}-e{\bm A} \ .
\end{equation}
The energy of the particle is determined by the kinematic momentum, $\epsilon=\epsilon_k$.
The absolute value of the kinematic momentum is equal to the Fermi momentum, $k=k_F$, where the Fermi
momentum  is determined by the condition $\epsilon_F=\epsilon_k$.
The cyclotron radius and the kinematic momentum are related by the Onsager relation
\begin{equation}
\label{rc1}
r_c=\frac{k}{e B} \ .
\end{equation}
Note that all the relations presented above are independent of the form of the dispersion $\epsilon_k$.
Hence our analysis is equally applicable to a semiconductor with quadratic dispersion, 
a semiconductor with nonquadratic dispersion, or to graphene.

Evaluation of the wave function phase along the arc trajectory is straightforward,
\begin{eqnarray}
\label{S}
&&\int{\bm k}\cdot d{\bm l}=kr_c(\pi-2\theta)\\
&&\int e {\bm A}\cdot d{\bm l}=-\frac{kr_c}{2}[\pi-2\theta-\sin2\theta]\nonumber\\
&&S=\int{\bm p}\cdot d{\bm l}=\int({\bm k}+e{\bm A})\cdot d{\bm l}
=\frac{k r_c}{2}[\pi-2\theta+\sin2\theta] \ .\nonumber
\end{eqnarray}
Here $\theta$ is the injection angle, see Fig.~\ref{Fig3}a.
Clearly this calculation assumes that
\begin{equation}
\label{nu}
\nu=k r_c \gg 1
\end{equation}
There is an additional contribution of $-\pi/2$ to the phase, due to the presence of a caustic,
as illustrated in panels b and c of Fig.~\ref{Fig3}.  
A caustic is a point where the 
trajectory is not uniquely defined, and there is a classical singularity in the 
intensity.
For a bundle of ``red'' trajectories, close to the negative solution of Eq.~\ref{rel1}, where
$\theta \to -|\theta| \pm \delta\theta$, $\delta \theta \ll |\theta|$, there is a crossing point
corresponding to the caustic, shown by the dashed lines in Fig.~\ref{Fig3}b.
Hence, there is a $-\pi/2$ addition to (\ref{S}), 
see e.g. Ref~\cite{LL2},
\begin{equation}
\label{S1}
S(-|\theta|) \to \frac{\nu}{2}[\pi+2|\theta|-\sin2|\theta|] - \pi/2
\end{equation}
On the other hand  a bundle composed of blue trajectories close to the positive solution of Eq.~\ref{rel1}, 
$\theta \to +|\theta| \pm \delta\theta$,  shown by dashed lines in Fig.~\ref{Fig3}c,
does not cross.  
Hence~\cite{LL2}
\begin{equation}
\label{S2}
S(+|\theta|) = \frac{\nu}{2}[\pi-2|\theta|+\sin2|\theta|] \ .
\end{equation}
Eqs. (\ref{S1}) and (\ref{S2}) determine phases of the semiclassical wave functions.
Finally, we note that a bundle of trajectories with an angular spread of $\delta \theta$ at the source will 
have a final spread at the detector of $\delta y$. This final spread can be determined via the Jacobian, 
\begin{equation}
\label{J}
J = \left|\frac{dL}{d\theta}\cos\theta\right|^{-1}=
\frac{1}{r_c|\sin2\theta|} \ .
\end{equation}
Multiplying by this to ensure conservation of flux, the Huygens kernel for particle propagation from source to detector
reads
\begin{eqnarray}
\label{Gsc1}
&&K(L,0) = a\sqrt{\frac{J}{2}}\left(e^{iS(-|\theta|)}+e^{iS(+|\theta|)}\right)\\
&=& e^{i\frac{\pi k \nu}{2}-i\frac{\pi}{4}}\sqrt{\frac{k}{\pi r_c|\sin2\theta|}} 
\sin \left[ \frac{ \nu}{2} \left( 2 |\theta|  - \sin2 |\theta| \right) + \frac{\pi}{4} \right]\ .\nonumber
\end{eqnarray}
Here $a$ is given by Eq.(\ref{aa}).
The normalization of (\ref{Gsc1}) comes from Eq.(\ref{H1}). The total probability flux calculated
with (\ref{H1}) near the source is $k/2$.
The total flux calculated with kernel (\ref{Gsc1}) is the same
\begin{equation}
\label{norm}
\int_0^{2r_c}|K(L,0)|^2\cos\theta dL =k/2 \ .
\end{equation}
We stress that (\ref{Gsc1}) is independent of the particle dispersion.

For small angles, $\theta \ll 1$, Eq.(\ref{Gsc1}) can be rewritten as
\begin{eqnarray}
\label{Gscs}
K(L,0) \approx e^{i\frac{\pi \nu}{2}-i\frac{\pi}{2}}
\sqrt{\frac{k}{2\pi r_c|\theta|}}
\sin\left[\frac{2}{3}\nu|\theta|^3+\pi/4\right] \ .
\end{eqnarray}
According to Eq.(\ref{rel1}) the angle $\theta$ is related  to the detuning from the edge of
the classical shade.
\begin{eqnarray}
\label{rel2}
\theta&\approx& \sqrt{\frac{-y}{r_c}}\nonumber\\
y&=&L-2 r_c \ .
\end{eqnarray}
This equation has meaning only for negative detuning, $y < 0$. Positive detuning, $y > 0$, 
corresponds to the classically forbidden region. In this ``shade" region, the intensity is zero.
Eq.(\ref{Gsc1}) rewritten in terms of 
\begin{equation}
\label{y}
\overline{y}=\frac{y}{r_c/\nu^{2/3}}
\end{equation}
reads
\begin{eqnarray}
\label{Gsc2}
K =  e^{i\frac{\pi (\nu-1)}{2}}\frac{\nu^{2/3}}{\sqrt{2}r_c}
\left\{
\frac{1}{\sqrt{\pi}|\overline{y}|^{1/4}}
\sin\left[\frac{2}{3}(-\overline{y})^{3/2}+\pi/4\right]
\right\} \ .
\end{eqnarray}
 The expression in curly brackets in Eq.(\ref{Gsc2})
 is the asymptotic expansion of the Airy function. 
Therefore, we conjecture that the general form of the Huygens kernel valid in both the
``bright" and ``dark" regions is
\begin{eqnarray}
\label{ai}
K(L) = e^{i\frac{\pi (\nu-1)}{2}}\frac{\nu^{2/3}}{\sqrt{2}r_c} Ai(\overline{y}) \ .
\end{eqnarray}
Eqs.(\ref{Gsc1}) and (\ref{ai}) represent the first major result of our work.
The phase factors in these Eqs. depend on the gauge, but the rest is gauge invariant.
We stress again that both Eqs. are independent of the particle dispersion
and equally applicable to semiconductors, graphene, etc.
According to (\ref{ai}) the typical spatial scale of the interference pattern in TMF is 
\begin{equation}
\label{dy}
\frac{\Delta y}{2r_c} \sim \frac{1}{\nu^{2/3}} \ .
\end{equation}\\

Eq.(\ref{ai}) assumes that the magnetic field is fixed but the distance between the source and
the detector is variable. In a typical experimental situation the distance is fixed, but the magnetic
field is variable. It is instructive to rewrite (\ref{ai}) in terms of the focusing field, $B_0$,
\begin{eqnarray}
\label{ai1}
K&=&e^{i\frac{\pi (\nu-1)}{2}}\frac{\nu^{2/3}}{\sqrt{2}r_c} Ai\left(\frac{2(B-B_0)}{B_0/\nu^{2/3}}\right) \\
B_0&=&\frac{2k}{eL} \ .\nonumber
\end{eqnarray}

In the derivation of (\ref{ai}) we made a logical leap from Eq.(\ref{Gsc2}) to Eq.(\ref{ai}).
We will now close the gap with a formal derivation of the Green's function for quadratic dispersion.
Note that while the Huygens' kernel is not equal to the Green's function, it is proportional to it.
The Hamiltonian and the spectrum reads 
\begin{eqnarray}
\label{h2}
&&H=\frac{{\bm \pi}^2}{2m}=\omega_c\left(a^{\dag}a+\frac{1}{2}\right)\nonumber\\
&&a=i\frac{\pi_-}{\sqrt{2B}}\ , \ \ \ a^{\dag}=-i\frac{\pi_-}{\sqrt{2B}}\nonumber\\
&&\epsilon_n=\omega_c(n+1/2) \ .
\end{eqnarray}
Here $\omega_c=eB/m$ is the cyclotron frequency.
Eigenstates in the gauge (\ref{gg}) are well known
\begin{eqnarray}
\label{eg1}
&&\psi=e^{ik_yy}\chi_n(x-x_0)\\
&&x_0=-\frac{k_y}{B} \ . \nonumber
\end{eqnarray}
Here $\chi_n$ are harmonic oscillator eigenfunctions.
Hence the Green's function for propagation of an electron/hole with energy $\epsilon\approx \epsilon_n$ from $y=0$ to $y=L$, see  Fig.\ref{Fig3}, during time $t=T/2$  is
\begin{eqnarray}
\label{G22}
G&=&\int\frac{d\epsilon}{2\pi}\sum_{k_y,n}e^{ik_yL}\frac{\chi_n^2(-x_0)}{\epsilon-\epsilon_0+i0}
e^{i\epsilon t}\nonumber\\
&\propto& e^{-i\epsilon_nT/2}\int \frac{dk_y}{2\pi}e^{ik_yL}\chi_n^2(-x_0)\ .
\end{eqnarray}
To evaluate the Green's function we use the semiclassical approximation for the oscillator
wave function
\begin{eqnarray}
\label{chi1}
\chi_n(x)&\approx& \frac{1}{p(x)}\cos S(x)\\
S(x)&=&\int_0^xp(x')dx'\nonumber\\
p(x)&=& \sqrt{2m\epsilon_n-m^2\omega_c^2x^2} \ . \nonumber
\end{eqnarray}
The $k_y$-integration in the propagator (\ref{G22}) is performed by the
stationary phase method
\begin{eqnarray}
\label{stat}
&&\int dk_y e^{ik_yL}\chi_n^2(-x_0)\propto \int dx e^{iBxL}\cos^2S(x)\\
&&\propto \int dx e^{iBxL}\cos^2S(x)\to \frac{1}{2}\int dx e^{iBxL} e^{-2iS(x)}\nonumber
\end{eqnarray}
At small $x' = x - x_0$, the wave function phase is
\begin{equation}
\label{s} 
S(x) =\int_0^{x'}p(x')dx'\approx kx'-\frac{B^2x'^3}{6k}=Br_cx'-\frac{Bx'^3}{6r_c} \nonumber
\end{equation}
and the integral at $x=0$ in (\ref{stat}) is transformed into
\begin{eqnarray}
\label{stat1}
&&\int dk_y e^{ik_yL}\chi_n^2(-x_0)\nonumber\\
&&\propto \int dx_0 \cos\left((BL-2Br_c)x_0+\frac{Bx_0^3}{3r_c}\right)\nonumber\\
&&=\int dx_0\cos\left(Byx_0+\frac{Bx_0^3}{3r_c}\right)\propto 
\int d\xi\cos\left({\overline y}\xi+\frac{\xi^3}{3}\right)\nonumber\\
&&{\overline y} =\frac{y}{r_c/(kr_c)^{2/3}}
\end{eqnarray}
The replaced integration variable in (\ref{stat1}) is $\xi=(kr_c)^{1/3}x_0/r_c$.
Expression in Eq.(\ref{stat1}) is the integral representation of the Airy function.
This calculation confirms that the Huygens' kernel (\ref{ai}) is proportional to the Airy
function.

\section{Spin-orbit coupled systems} 
\label{withspinorbit}

In this section we consider SOIs with given winding numbers. We can express the Hamiltonian as~\cite{Bladwell}
\begin{eqnarray}
{\cal H}_{SOI} = i \frac{\gamma_n}{2} k_-^n \sigma_+ +h. c.
\label{spinorbit1}
\end{eqnarray}
Here n is the winding number, $k$ is the particle momentum, $k_{\pm}=k_x\pm i k_y$, and
$\sigma_\pm = \sigma_x \pm i \sigma_y$, are the Pauli matrices describing 
the effective spin 1/2. 
For electron systems, these matrices represent the electron spin, while for 
holes which have internal angular momentum  $J=3/2$, the  matrices describe
the two level heavy hole subsystem $J_z=\pm 3/2$.
We consider the cases $n=1$, the linear Rashba interaction~\cite{Bychkov1984}; and 
$n=3$, the qubic Rashba interaction. The coefficients $\gamma_n$ in both cases are real.

The SOI is equivalent to a momentum dependent effective Zeeman
{\it in-plane} magnetic field, ${\cal B}({\bf k})$. 
\begin{eqnarray}
\label{sp1}
{\cal H}_{SOI} = -{\cal B}_{\bm k}\cdot {\bm \sigma} \ .
\end{eqnarray}
For a particle moving along a circular trajectory Eq.(\ref{spinorbit1}) results in the following effective magnetic field
\begin{eqnarray}
\label{spinorbit11}
&&{\cal B}_{\bm k}=\gamma_n k^n(-\sin n\varphi,\cos n\varphi,0) \\
&&{\bf k} = k(\cos \varphi, \sin \varphi,0) \ ,\nonumber
\end{eqnarray}
where $\varphi$ is the axial angle in $k$-space. 
Here we consider only the case of sufficiently strong SOI, $|{\cal B}|\gg \omega_c$,
so the spin dynamics are adiabatic, spin is always parallel or antiparallel to the 
local direction of the effective  magnetic field, ${\cal B}({\bf k})$. 
The physical meaning of the winding number is evident in this limit,
it is the number of spin rotations for one full revolution of the particle in a magnetic field.

In the adiabatic approximation the orbital dynamics of the particle are described by the
Hamiltonian~\cite{Zulicke2007,Bladwell}
\begin{equation}
\label{Had}
H=\epsilon_k\pm |{\cal B}_{\bf k}| \ ,
\end{equation}
where $\epsilon_k$ is spin independent part of the dispersion.
We remind that $k$ is given by Eq.(\ref{kp}).
Eq.(\ref{Had}) results in classical double focusing. The separation between two classical
focusing points is~\cite{Bladwell}
\begin{equation}
\label{cl}
\frac{\Delta L}{2r_c}\approx \frac{\gamma_nk_F^n}{\epsilon_F} \ .
\end{equation}
It is instructive to estimate the ratio of the classical separation to the distance
between the interference fringes (\ref{dy}).
\begin{equation}
\label{rat1}
\frac{\Delta L}{\Delta y}\sim \frac{\gamma_nk_F^n}{\epsilon_F}\ \nu^{2/3} \sim 
\frac{|{\cal B}|/\omega_c}{2\nu^{1/3}}\ .
\end{equation}
Even within the validity of the spin-adiabatic approximation, $|{\cal B}|/\omega_c \gg 1$,
the ratio (\ref{rat1}) can be about unity, $\frac{\Delta L}{\Delta y} \sim 1$,
the quantum and the classical scales are comparable.
Nevertheless in this work for simplicity we consider the case $\frac{\Delta L}{\Delta y} \gg 1$,
the SOI is so strong that there are two well separated classical peaks with interference 
fringes on each peak.

In the adiabatic limit there are two Fermi surfaces determined by the equation
\begin{equation}
\label{fs}
\epsilon_F=\epsilon_k\pm |{\cal B}_{\bf k}|
\end{equation}
We assume that $\epsilon_k$ is independent of the direction of ${\bm k}$,
hence Eq.(\ref{fs}) results in two circular Fermi surfaces with Fermi momenta $k_{\pm}$.
Corresponding cyclotron radii are
\begin{eqnarray}
r_{c, \pm} = \frac{k_\pm}{e B}
\end{eqnarray}
In typical experimental systems, $\epsilon_k$ is not quite isotropic, higher order corrections
lead to anisotropies. 
In this work we neglect the dispersion anisotropy effects.

\begin{figure}[h!]
     {\includegraphics[width=0.25\textwidth]{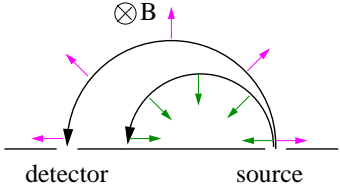}}
	\caption{Two spin-orbit split classical trajectories. The coloured arrows show the spin.
This is the case of the winding number $n=1$.}
\label{Fig1s}
\end{figure}

\begin{figure}[h!]
     {\includegraphics[width=0.2\textwidth]{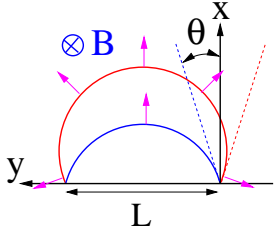}}
	\caption{Two interfering trajectories corresponding to $\epsilon_F=\epsilon_k- |{\cal B}_{\bf k}|$
 The colored arrows show the spin. To avoid ``overcrowding'' on the blue trajectory we show
spin only at one point.
This is the case of the winding number $n=1$.}
\label{FigB}
\end{figure}

There are two classical trajectories focused at different points.
In Fig.~\ref{Fig1s} the spin of these trajectories is indicated by coloured arrows, the picture
corresponds to the winding number $n=1$.
We consider the interference picture near one of the classical peaks, 
$k=k_+$, $r_c=r_{c,+}$ (the trajectory with ``magenta'' spin in Fig.~\ref{Fig1s}).
The analysis of Section \ref{nospinorbit} is fully applicable
in this case. However, we must also account for the effects of spin dynamics.
To illustrate the spin dynamics in  Fig.~\ref{FigB}  we redraw Fig.\ref{Fig3}a with two interfering
trajectories, but now present the spin-orbit coupled case with $n=1$.
There are two effects related to the spin, (i) projection of the source/detector spin state on the
spin eigenstate of the trajectory, (ii) Berry phase accumulated along the trajectory.

The spin eigenstate of the Hamiltonian ({\ref{spinorbit1}),(\ref{sp1}) reads
\begin{eqnarray}
\label{ses}
\chi = 
\begin{pmatrix} 
-i s  e^{-i n\varphi/2} \\
e^{in \varphi/2}
\end{pmatrix} \ ,
\end{eqnarray}
where $\varphi$ varies along the trajectory $\pi-\theta > \varphi > \theta$, see Eq.(\ref{spinorbit11}) and Fig.~\ref{FigB}.
Here $s=\pm 1$, the energy is $\epsilon=\epsilon_k\mp |{\cal B}_{\bf k}|$, so Fig.~\ref{FigB} corresponds to $s=+1$.
Hence the eigenspinor at the source is
\begin{eqnarray}
\label{sorse}
\chi_{i} = \frac{1}{\sqrt{2}}
\begin{pmatrix} 
-i s  e^{-i n\theta/2} \\
e^{in \theta/2}
\end{pmatrix} \ ,
\end{eqnarray}
and the eigenspinor at the detector is
\begin{eqnarray}
\label{det}
\chi_{f} = \frac{1}{\sqrt{2}}
\begin{pmatrix} 
-i s  e^{-i n(\pi-\theta)/2} \\
e^{in (\pi -\theta)/2}
\end{pmatrix} = \frac{1}{\sqrt{2}}
e^{-in\pi/2}
\begin{pmatrix} 
-is  e^{i n \theta/2} \\
- e^{-in\theta/2}
\end{pmatrix} \ .
\end{eqnarray}
The Huygens' kernel is determined by taking the outer
product of the initial and final eigenspinors, $\chi_f\chi_i^{\dag}$,
and hence is a matrix, with Eq.(\ref{Gsc1}) replaced by
\begin{widetext}
 \begin{eqnarray}
\label{kso}
 K^s_{fi} =\frac{1}{2} e^{i\frac{\pi (\nu_s-1-n)}{2}}
\sqrt{\frac{k_{s}}{\pi r_{cs}|\sin 2\theta|}}
\begin{pmatrix}
  \sin \left[ \frac{ \nu}{2} \left( 2 |\theta|  - \sin2 |\theta| \right) -n|\theta|+ \frac{\pi}{4} \right]
&-i s 
\sin \left[ \frac{ \nu}{2} \left( 2 |\theta|  - \sin2 |\theta| \right) + \frac{\pi}{4} \right] \\
-i s  \sin \left[ \frac{ \nu}{2} \left( 2 |\theta|  - \sin2 |\theta| \right) + \frac{\pi}{4} \right]
& -   \sin \left[ \frac{ \nu}{2} \left( 2 |\theta|  - \sin2 |\theta| \right) +n|\theta|+ \frac{\pi}{4} \right]
 &\\
 \end{pmatrix}
 \end{eqnarray}
Naturally the kernel depends on the index $s=\pm 1$ enumerating classical trajectories.
The semiclassical parameter $\nu$ is defined similarly to Eq.(\ref{nu}), $\nu_s= k_s r_{cs}$.
Finally, using the same logic as in Section \ref{nospinorbit} we rewrite (\ref{kso}) as
 \begin{eqnarray}
\label{kso1}
 K^s_{fi} &=&e^{i\frac{\pi (\nu_s-1-n)}{2}}\frac{\nu_s^{2/3}}{2\sqrt{2}r_{cs}}
\begin{pmatrix}
 Ai(\overline{y}_s+n/\nu_s^{1/3}) 
&-i s Ai(\overline{y}_s)\\
-i s  Ai(\overline{y}_s)
&-  Ai(\overline{y}_s-n/\nu_s^{1/3}) &\\
 \end{pmatrix}\nonumber\\
&=&e^{i\frac{\pi (\nu_s-1-n)}{2}}\frac{\nu_s^{2/3}}{2\sqrt{2}r_{cs}}\left[(\sigma_z-is \sigma_x)Ai(\overline{y}_s)
+\frac{n}{\nu_s^{1/3}}Ai^{\prime}(\overline{y}_s)\right] \ .
 \end{eqnarray}
\end{widetext}
Here ${\overline y}_s$ is defined by Eq.(\ref{y}) with $r_c \to r_{cs}$, $\nu \to \nu_s$;
$Ai^{\prime}$ is derivative of the Airy function.
The Berry phase $e^{-in\pi/2}$ in the prefactor and the spin structure $(\sigma_z-is \sigma_x)$
in the leading semiclassical term are very simple, they immediately follow from Fig.\ref{Fig1s}
without any calculation. The subleading semiclassical term $\sim n/\nu_s^{3/2}$ is enhanced by
the winding number, thus it is especially important for $n=3$.
In a realistic experiment $\nu$ cannot be large, $\nu \lesssim 100$, hence  $3/\nu^{1/3} > 0.6$.

\section{Discussion}
\label{disc}

Equations (\ref{HF}), (\ref{Gsc1}), (\ref{ai}), and (\ref{kso1}) solve the problem of the quantum interference
in transverse magnetic focusing. If the source and the detector are
quantum point contacts, they can be modelled by standing waves in the y-direction~\cite{Bladwell}.
\begin{eqnarray}
\label{psisd}
\psi_{s} &=& \chi_s \sin\left(\frac{\pi y}{W} \right)  \ \ \ \ \ \ \  \ \ \ \ \  0 < y < W\nonumber\\
\psi_{d} &=& \chi_d \sin\left(\frac{\pi (y-L)}{W} \right)  \ \ \  L < y < L+W \ .
\end{eqnarray}
Here $W$ is the width of the channel and $\chi_s$, $\chi_d$ are spin functions of the source/detector.
Evidently the aperture cannot be smaller than half of the wave length, $W > \lambda/2$.
According to the Huygens' principle the signal observed in the TMF experiment is
\begin{eqnarray} 
\label{int}
I \propto \sum_{s=\pm 1}\left|\int \psi_{d}^{\dag}(y_2) K^s(y_2-y_1) \psi_{s}(y_1)dy_1dy_2\right|^2\ .
\end{eqnarray}
The integration is performed over the apertures of the source and the detector.

Let us consider two examples with experimentally reasonable parameters.
First we address the case without SOI. Let the Fermi momentum be $k=0.1nm^{-1}$, which
corresponds to the wavelength $\lambda \approx 63nm$ and the number density $n=1.6 \ 10^{11}cm^{-2}$. 
The distance between the source and the detector is $L=2000nm$, and the corresponding
value of the magnetic field at the  classical edge of the bright region is $B_0=66mT$.
Interference patterns obtained with Eqs.(\ref{int}),(\ref{psisd}) and
apertures $W=30nm\approx \lambda/2$ and $W=60nm\approx \lambda$ are plotted in Fig.~\ref{FigNoSOI}.
\begin{figure}[h!]
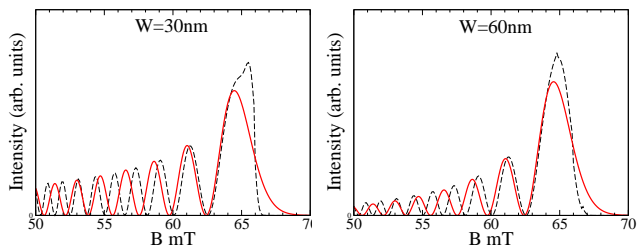

     {\includegraphics[width=0.23\textwidth]{FigW30NSOI.eps}}
     {\includegraphics[width=0.23\textwidth]{FigW60NSOI.eps}}
	\caption{Interference patterns versus focusing magnetic field for
two values of the quantum point contact apertures $W$. The  spin orbit interaction is zero
and the separation between the source and the detector is 2000nm.
 The black dashed curves are obtained with the semiclassical kernel 
(\ref{Gsc1}) and the red curves are obtained with the full ``Airy'' kernel (\ref{ai1}).
}
\label{FigNoSOI}
\end{figure}
%The black dashed curves  are obtained with the semiclassical kernel (\ref{Gsc1}) and the red
%solid curves are obtained with the ``Airy'' kernel (\ref{ai1}).
We note that the ``Airy'' kernel is perfectly accurate to describe the transition from the ``bright'' 
region to the ``dark'' one. The semiclassical approximation incorrectly transfers some spectral weight
from the ``dark'' region to the first bright fringe.
On the other hand the ``Airy'' approximation fails deeper in the ``bright'' region because 
the condition,  $\theta \ll 1$, is not valid in this region.
 At the same time the  semiclassical approximation is perfectly valid in the ``bright" region. 
The difference becomes significant by the 3rd-4th interference fringe.

Now we turn to the case of strong SOI with the winding number $n=3$.
Let the smaller Fermi momentum be $k_-=0.1nm^{-1}$ ($\lambda \approx 63nm$)
and the larger Fermi momentum $k_+=0.14nm^{-1}$ ($\lambda \approx 45nm$).
The corresponding  density is $n=2.4 \ 10^{11}cm^{-2}$. 
The distance between the source and the detector is $L=2000nm$.
The corresponding magnetic field at the  classical edge of the bright region for $k_-$
is $B_0=66mT$, while for $k_+$, $B_{0-}=92mT$.
In our calculation we assume that the source produces unpolarized holes,
so in Eq.(\ref{int}) we average over polarizations. 
Interference patterns obtained with Eqs.(\ref{int}),(\ref{psisd}) and
apertures $W=30nm$ and $W=60nm$ are plotted in Fig.~\ref{FigSOI}.
\begin{figure}[h!]
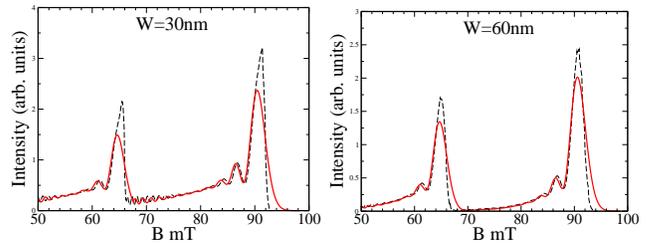

     {\includegraphics[width=0.23\textwidth]{FigW30SOI.eps}}
     {\includegraphics[width=0.23\textwidth]{FigW60SOI.eps}}
	\caption{Interference patterns versus focusing magnetic field for
two values of the quantum point contact apertures $W$. The  spin orbit interaction is 
so strong that there are two spin orbit split focusing peaks.
The separation between the source and the detector is 2000nm.
 The black dashed curves are obtained with the semiclassical kernel 
(\ref{kso}) and the red curves are obtained with the full ``Airy'' kernel (\ref{kso1}).
{\cred This figure corresponds to the unpolarized source/detector}
}
\label{FigSOI}
\end{figure}
Similar to the case without  SOI the semiclassical approximation incorrectly transfers some 
spectral weight from the ``dark'' region to the first bright fringe. 
It is worth noting that the smearing of the interference picture in Fig.\ref{FigSOI} compared
to Fig.\ref{FigNoSOI} is due to the spin phase factor contribution $n/\nu^{1/3}$ in the
arguments of the Airy functions in Eq.(\ref{kso1}). The smearing is most significant for
unpolarized holes. 
The assumption of fully unpolarized source/detector in the case
of a strong Rashba interaction is not well justified.
As an alternative limit one can  consider the source/detector as fully polarized
by the Rashba interaction. In this case the intensity is proportional to the sum of the 
matrix elements
of the matrix (\ref{kso1}) with appropriate coefficients.
Plots of the intensity in this case are presented in Fig.\ref{FigSOIP}.
\begin{figure}[h!]
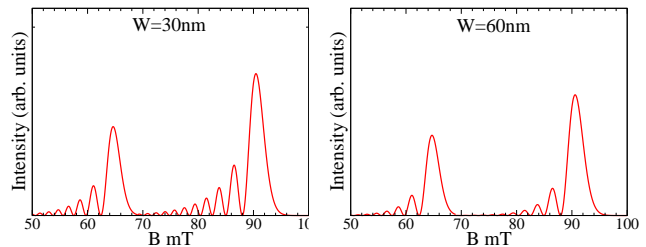

     {\includegraphics[width=0.23\textwidth]{FigW30SOIP.eps}}
     {\includegraphics[width=0.23\textwidth]{FigW60SOIP.eps}}
	\caption{Interference patterns versus focusing magnetic field for
two values of the quantum point contact apertures $W$. The  spin orbit interaction is 
so strong that there are two spin orbit split focusing peaks.
The separation between the source and the detector is 2000nm.
 This figure corresponds to a fully polarized source/detector.
}
\label{FigSOIP}
\end{figure}
The interference picture is sensitive to polarizations, and exploration 
of interference from polarized QPCs could prove experimentally interesting.

In conclusion, we have derived the Huygens' kernels for the Huygens principle in
transverse magnetic focusing of electrons/holes in two dimensional heterostructures.
The derived technique reduces the focusing problem to evaluations of a simple integral.
The developed technique is very general and can be applied to a system with practically any kind
of spin orbit interaction with cylindrical symmetry.
Specifically we have considered the  case of no spin orbit interaction,
and of strong spin orbit interactions of Rashba type with given
spin winding number.

\section{Acknowledgements}
We thank Alexander Milstein, Alex Hamilton, Scott Liles, Matthew Rendell, Ashwin Srinivasan, Uli Zuelicke,
Tommy Li,  Dmitry Miserev,  and Yaroslav Kharkov for important  stimulating discussions. 
The work has been supported by the Australian Research Council grant DP160103630.

\end{document}